\documentclass{emulateapj}

\newcommand{\GA}{{\it Great Attractor}}

\newcommand{\HI}{H\,{\sc i}}
\newcommand{\HII}{H\,{\sc ii}}

\newcommand{\cm}{~cm$^{-2}$}
\newcommand{\kms}{~km\,s$^{-1}$}
\newcommand{\mjysr}{mJy\,sr$^{-1}$}
\newcommand{\kkms}{km\,s$^{-1}$}
\newcommand{\vhel}{$v_{\rm hel}$}

\newcommand{\wtw}{$w_{\rm 20}$}
\newcommand{\wfi}{$w_{\rm 50}$}

\newcommand{\FHI}{$F_{\rm HI}$}

\newcommand{\MHI}{$M_{\rm HI}$}

\newcommand{\Msun}{~M$_{\odot}$}
\newcommand{\LLsun}{L$_{\odot}$}

\newcommand{\Ho}{H$_{\rm 0}$}

\newcommand{\AV}{$A_{\rm V}$}

\slugcomment{Jarrett et al. 2007, AJ, 133, no. 3}

\shorttitle{Galaxies Within the Great Attractor}
\shortauthors{Jarrett et al.}

\begin{document}


\title{Discovery of Two Galaxies Deeply Embedded in the Great Attractor Wall}


\author{T.H. Jarrett}
\affil{Spitzer Science Center, IPAC 100-22, Caltech, Pasadena, CA 91125}
\email{jarrett@ipac.caltech.edu}

\author{B.S. Koribalski}
\affil{Australia Telescope National Facility, CSIRO, PO Box 76, Epping, 
       NSW 1710, AU}

\author{R.C. Kraan-Korteweg \& P.A. Woudt}
\affil{Dept. of Astronomy, University of Cape Town, Private Bag X3, 
       Rondebosch 7701, RSA}

\author{B.A. Whitney}
\affil{Space Science Institute, 4750 Walnut Street, Suite 205, Boulder, 
       CO 80301}

\author{M.R. Meade, B. Babler  \& E. Churchwell}
\affil{Astronomy Dept,, 475 N. Charter St.,
University of Wisconsin-Madison, Madison, WI 53706}

\author{R.A. Benjamin}
\affil{Physics Dept,, 800 W. Main St., University of Wisconsin-Whitewater,
Whitewater, WI 53190}

\and

\author{R. Indebetouw}
\affil{Astronomy Dept,, PO 318, University of Virginia,
Charlottesville, VA  22903-0818} 



\begin{abstract}

We report on the discovery of two spiral galaxies located behind 
the southern Milky Way, within the least explored region
of the \GA. They lie at ($\ell,b \approx
317\degr, -0.5\degr$), where obscuration from Milky Way stars and
dust exceeds 13 to 15 mag of visual extinction.  The galaxies were the most
prominent of a set identified using mid-infrared images of the low-latitude
($|b| < 1\degr$) Spitzer Legacy program GLIMPSE.  Follow-up \HI\ radio
observations reveal that both galaxies have redshifts that place them
squarely in the Norma Wall of galaxies, which appears to extend
diagonally across the Galactic Plane from Norma in the south to
Centaurus/Vela in the north. We
report on the near-infrared, mid-infrared and radio properties of these newly
discovered galaxies, and  
discuss their context
in the larger view of the \GA. 
The work presented here demonstrates that
mid-infrared surveys
open up a new window to study galaxies in the Zone of Avoidance.

\end{abstract}





\section{Introduction}



Our emerging portrait of the universe is that of an intricate
cosmic web of galaxies arrayed in long filaments, sheets and bubbles that
intersect to form dense mass concentrations.
The key outstanding problem is the
distribution and nature of dark matter and dark energy that drives the
dynamics of the expanding cosmos.  The study of the local Universe,
including its peculiar motions and its clustering at the largest
size scales, is essential for
structure formation in the early Universe and its relation to the
formation and subsequent evolution of galaxies.  
Studying superclusters both near
and far is paramount in decoding the mass density of the universe.

We now recognize that the core of one of the most important mass
concentrations in the local universe is located behind the southern
Milky Way ($\ell \sim 320\degr, b \sim 0\degr$).  The so-called
{\it Great Attractor} (GA) is 
pulling on the 
Milky Way and the Local Group of galaxies, its gravitational influence stretching
beyond Virgo and the Local Supercluster of galaxies
(Lynden-Bell et al. 1988; Burstein et al. 1990;
Kocevski \& Ebeling 2006; Erdogdu et al. 2006a,b). 
The search for the elusive GA covers a very large
expanse of sky that is within the Zone of Avoidance (ZoA), where
traditional surveys are limited by the foreground Milky Way.
It has been attributed to or associated with visible galaxy overdensities
in Hydra, Centaurus, Pavo and Indus (e.g. Lahav 1987; Lynden-Bell, Lahav, \& Burstein 1989),
and most recently with dedicated surveys of the ZoA,
in Norma and Vela (e.g., see Kraan-Korteweg \& Lahav
2000, Kraan-Korteweg 2005, for reviews).  

To date, the most prominent density peak to be discovered in
the southern ZoA is the Norma cluster (Abell 3627; Abell, Corwin \&
Olowin 1989), whose cumulative mass and richness is comparable to the Coma
cluster (Kraan-Korteweg et al.~1996; Woudt, Kraan-Korteweg \& Fairall
1999).  It is located below the southern Galactic equator (325\degr,
$-7$\degr, 4900\kms), the limit where optical and
near-infrared surveys become heavily incomplete. 
To the north, emerging on the other side
of the darkest region of ZoA are the Cen-Crux and CIZAJ1324.7--5736
clusters, both at similar recessional velocities as Norma (Radburn-Smith et al.~2006).
For a velocity range between 4000 and 6000\kms, a continuous line of
galaxies appear to extend northward from Pavo-Indus up through Norma, bending to
lower longitudes and joining with Cen-Crux/CIZAJ1324.7--5736 in the
northern Galactic hemisphere (see Fig.~1 in Jarrett 2004 for a 3D
view of these large scale structures).
This large filamentary structure, named the "Norma Wall",
runs diagonally behind the southern Milky
Way (Kraan-Korteweg et al. 1994; Kraan-Korteweg 2005),
severely hampering the construction of sufficiently
detailed large-scale flow maps that are needed to identify the full extent,
shape and mass of the GA. Despite dedicated multi-wavelength efforts
in mapping the GA in the last decade (Kraan-Korteweg 2005),
a complete census of the
density and total mass is still unrealized.

From above the Earth's blocking atmosphere, the mid-infrared (MIR) 
opens a new and promising window for detecting ZoA
galaxies. The MIR can penetrate the thick layer of Galactic gas and dust, whilst
being sensitive to both stellar photospheric emission of early-type
galaxies and interstellar emission from star-forming late-type disk
galaxies.  The Spitzer Legacy project ``Galactic Legacy Infrared
Mid-Plane Survey Extraordinaire'' (GLIMPSE; Benjamin et al. 2003), 
surveyed a large fraction of the Milky Way (within $|b| < 1$\degr) 
using the IRAC camera ($3.6-8 \micron$), passing across the most 
opaque part of the GA.  GLIMPSE provides a readily available  
data set to test the viability
of probing the ZoA with MIR imaging.

Inspection of the GLIMPSE MIR images revealed a number of galaxy candidates
peaking through the formidable foreground of Galactic dust, gas and stellar
confusion. Intriguingly, a concentration of galaxy candidates appear to be projected
against the Norma Wall exactly where the wall supposedly bends
from the southern hemisphere to the northern hemisphere.  To confirm
their extragalactic nature, we have followed-up two of the most
promising galaxy candidates using deep near-infrared (NIR) imaging
(Sect.~2.2) and radio observations (Sect.~2.3).  Both were found to
have radial velocities that are consistent with membership in the
Norma Wall (Sect.~3.2), thus providing compelling evidence that an extragalactic 
bridge exists
between the southern and northern halves of the Plane.  In Section 3 we
report on the properties of these ZoA galaxies and in Section 4 discuss their context
in the larger view of the GA.

\section{Observations \& Data Reductions}


\subsection{Spitzer Mid-Infrared Observations}
Active study of the Galactic Plane is now underway with the Spitzer Space Telescope.
The heart of the GA has been partially mapped by GLIMPSE using the 
Spitzer Infrared Array Camera (IRAC), covering the window between
$3.6-8 \micron$.
The relatively shallow GLIMPSE images
are comprised of two epochs of two second exposures, typically achieving 
a $5\sigma$ surface brightness
sensitivity of 2.7, 2.2, 6.3 and 5.6 \mjysr\ for IRAC 3.6, 4.5, 5.8
and 8.0 $\micron$, respectively.  
The Spitzer/GLIMPSE data reduction
pipeline creates astrometric and photometric absolute-calibrated
images that are combined into a single mosaic per band.  The composite 
point source FWHM is $\sim$2.5\arcsec.

At longer wavelengths, the MIPSGAL survey (Carey et al. 2006)
covered the same GLIMPSE footprint of the Milky Way using the
Spitzer Multiband Imaging Photometer for Spitzer (MIPS).
The diffraction limit at $24\micron$ is approximately 6 arcsec,
roughly a factor of two larger than the IRAC imaging,
adequate for measuring the dust-emitting continuum
from the GLIMPSE galaxies.
A combined IRAC + MIPS panorama of the GA is
shown in Fig.~1, simultaneously illustrating 
the incredible beauty of the Milky Way and the major challenge
we are faced as we look behind the Galactic veil.

Visually scanning the set of GLIMPSE images that crossed the GA, we
identified a number of resolved sources to be promising 
extragalactic candidates for follow-up
study. The two
most prominent ones (see Fig.~1) lie at $\alpha,\delta$(J2000) =
$14^{\rm h}48^{\rm m}36.0^{\rm s},-60\degr07\arcmin15\arcsec$, ($l,b$
= 317.04\degr, -0.50\degr) and $14^{\rm h}47^{\rm m}45.1^{\rm s},
-60\degr17\arcmin04\arcsec$ ($l,b$ = 316.87\degr, -0.60\degr), and are 
referred to as G\,1 and  G\,2, respectively.  
These sources were found in relatively low $8\micron$ and $24\micron$ background regions
compared to the very bright \HII\ emission encircling the bubble-like clearing.
The stellar confusion noise and the visual
extinction, however, were extremely high, rendering both galaxies near
the detection limit of GLIMPSE.  

Photometry was extracted
from the four IRAC bands and the MIPS $24\micron$ image
using a set of nested elliptical apertures
centered on each galaxy, where the images were first cleaned of
contaminating, foreground stars.  The background was determined using
an annulus that was well outside of the galaxy light, but still local
to the galaxy environment and considerably smaller than the scale of
the background ISM gradients.  For the IRAC measurements, 
the flux density of each source was aperture corrected using the "extended
source" prescription recommended by the Spitzer Science Center
\footnote{http://ssc.spitzer.caltech.edu/irac/calib/ExtSrcCal/}.
The expected uncertainty in the IRAC photometry
(measurement errors plus calibration) is $\sim$10\%.
For the MIPS $24\micron$ measurements, the appropriate aperture
correction was determined to be $\sim$1.1 (see Figure 3.2 of the MIPS data
handbook), while another factor of 1.04 was applied to the integrated
fluxes to account for the color correction between normal galaxies
and the MIPS absolute calibration standard (see Table 3.12 in the MIPS
data handbook).  The expected uncertainty in the MIPS photometry
is $\sim$10-15\%.

\subsection{IRSF Near-Infrared Observations} 
Near-infrared $JHK_s$ images of the region were obtained with the {\sc
sirius} camera on the 1.4m Infrared Survey Telescope (IRSF; Nagata \&
Glass 2000, Nagayama et al. 2003) in South Africa during April 4/5,
2006.  The exposures were relatively long, reaching the stellar
confusion limit at $2\micron$ for much of the region studied.  
The
images were reduced by subtracting median "sky" images, then
combined with astrometric and photometric calibration using
2MASS stars in the field.  The resulting mosaics, with
$\sim$0.9\arcsec\ FWHM profiles, were deep enough to detect the two
GLIMPSE galaxies in all three NIR bands (see Fig.~1 inserts for
combined NIR and MIR images).  Photometry was extracted using the same
set of elliptical apertures and methodology devised for the MIR images.  The largest
aperture corresponded to an isophotal surface brightness of 20.6 mag/arcsec$^2$.

\subsection{ATCA \HI\ Observations} 

\HI\ synthesis observations were obtained with the Australia Telescope
Compact Array (ATCA)
in the EW352 configuration on April 19/20, 2006. The pointing position 
was $\alpha,\delta$(J2000) = $14^{\rm h}48^{\rm m}36^{\rm s}, 
-60\degr\,06\arcmin\,00\arcsec$ ($\ell,b$ = 317.05\degr, --0.48\degr). 
Since the galaxy redshifts were unknown but a connection with the GA 
anticipated, we observed with the large bandwidth of 16 MHz centered on 
1397 MHz, leading to a velocity coverage ranging from $\sim$3500\kms\ 
to 6000\kms. With 512 channels, the channel width resulted in 6.7\kms. 
The field of view (primary beam FWHM) at 1397 MHz is 34.1\arcmin. The 
total observing time was $\sim$12 hours. 
We observed PKS\,0823--500 and 
PKS\,1934--638 for bandpass and flux calibration. The phase calibrator 
PKS\,1338--58 was observed for 3 minutes every 40 minutes.

Data reduction was carried out with the {\sc miriad} software package
using standard procedures. After identifying the velocity range of the
\HI\ signal we used the line-free channels to fit and subtract the
radio continuum emission. Because of a strong confusing radio source,
\HII\ region PMN\,J1445--5949, $\sim$30\arcmin\ from the pointing
center (Fig.~1, upper right), we used a 3rd order polynomial fit in
{\it uvlin} to subtract the continuum emission. After Fourier-transformation
of the \HI\ channel maps using `natural' weighting, the images were
cleaned and restored with a synthesized beam of $144\arcsec \times 
114\arcsec$. The measured r.m.s. noise per 20\kms\ wide channel is 
$\sim$2 mJy\,beam$^{-1}$. 

\section{Results}

\subsection{Infrared Emission and Foreground Extinction}
The dust-obscured GLIMPSE galaxies are only visible at wavelengths
longward of $1\micron$, significantly reddened due to selective
extinction.  We estimate the Galactic extinction using three 
independent techniques: dust column density, stellar colors
and galaxy colors.  (In the next section, 3.2, we estimate the foreground
extinction using the radio and photometric properties combined with the 
Tully-Fisher relation.) 
The goal is to arrive at a consistent
value of the extinction that arises from the edge-on disk of the Milky Way.
An accurate extinction correction will enable full characterization
of the ZoA galaxies.

Far-infrared emission, as measured by the IRAS and COBE space telescopes, 
can be used as an effective proxy of
the dust column density of the Milky Way.
The average dust extinction as
inferred from the IRAS/DIRBE far-infrared maps (Schlegel et al. 1998) is \AV\
$\sim 17.2$ and 17.7 mag, for G\,1 and G\,2 respectively.   
The accuracy of these estimates is limited by
the angular resolution and absolute calibration of
the telescopes and detectors, as well as confusion from
Milky Way stars that are not associated with the interstellar 
dust column. Moreover, studies suggest that the FIR-inferred
extinctions tend to be overestimated when the dust column density is
large (e.g., molecular clouds; see, e.g., Arce \& Goodman, 1999;
Nagayama et al. 2004;
Schr\"oder et al. 2006).

Shifting to the stellar population of the Galactic Plane,
field stars can be exploited
at shorter wavelengths to infer the extinction
by analyzing
the red giant branch of the
stellar color-magnitude diagram (CMD).
Accordingly, 2MASS near-infrared photometry of stars
in the direction of the GA is used to construct
the CMD; 
recent examples
using this method include
Ferraro et al. (2006),
Nishiyama et al. (2006),
Rocha-Pinto et al. (2006),
L{\'o}pez-Corredoira et al. (2005),
and Salaris \& Girardi (2005).
The 2MASS JHKs CMD towards (l=317\arcdeg, b=-0.6\arcdeg, radius=0.3\arcdeg)
clearly separates into a distinct dwarf and red clump sequence as well
as a strong bright red tail of reddened giants.
The red clump extends to a distance of 5.5$\pm$0.5 kpc and
$A_V$=8$\pm$1 mag using red clump intrinsic properties K=-1.5 mag and J-K=1.3 mag
(Bonatto et al. 2004, Salaris \& Girardi 2002; and
using $A_V/A_K$=8.8, Draine 2003).  This
likely corresponds to the Scutum-Crux arm at about that distance and
direction; however, the red clump is unlikely to be able to trace the full
extent of the disk at 2MASS resolution and sensitivity.
The bright giants are the only population that can reliably show the
full extent of the Galactic disk at 2MASS sensitivity and resolution.
Assuming constant NIR color for that population reveals a red giant
branch extending to a dereddened K magnitude of $\simeq$11.75 mag, and a
maximum (95$^{th}$ percentile) extinction $A_V$=14$\pm$1 mag.
So between the FIR emission and the colors of field giants, the
foreground extinction is estimated to be 14 - 17 mag in $A_V$.

Finally, the galaxy colors themselves can be used to
estimate the foreground extinction. 
For galaxies in the local
universe, the NIR window is sensitive
to the old stellar population, regardless of Hubble type, 
whose photospheric light
dominates the spectral energy distribution (SED) between
1 and $2 \micron$.  Consequently, the
NIR colors of all types of galaxies are
nearly identical (e.g., Jarrett 2000).  
Comparing the observed NIR colors with
the expected colors yields an estimate of the color reddening
and hence, line-of-sight extinction.  Accordingly, we have constructed
the galaxy SEDs from the NIR and MIR measurements,
which are directly compared with the expected light distribution of an
old galaxy ({\it t} = 13 Gyrs) constructed from the GRASIL population
synthesis (Silva et al. 1998).  The model galaxy is weighted heavily to fit
the peak in the SED light between $1 - 2\micron$,
with lower weighting applied to the window between
$3 - 5\micron$ (where the presence of hot dust continuum and star formation
line emission
are beginning to become important).  Extinction is a free parameter
in the fit.
The resultant best fit SED 
corresponds to a total visual
extinction (with $A_V/A_K$=8.8) of $A_V =$ 15.1 and 13.0 $\pm$ 0.2 mag 
for the respective galaxies.
We can estimate the internal extinction contribution
using the measured axis ratio and the prescription from
Masters, Giovanelli \& Haynes (2003): $A_k = 0.26\times
log_{10}(a/b)$, where b/a is the disk axis ratio.  
The axis ratios as measured with the deep NIR images (Table~1)
suggest an  internal visual extinction of
$\sim 1.4$ mag for  G\,1, and 0.6 mag for  G\,2 at visual wavelengths.
Hence, the total foreground extinction follows:
$A_V =$ 13.7 and 12.4 $\pm$ 0.2 mag
for the respective galaxies.
These SED-derived results are in good agreement with the 
CMD-derived values, and are within $\sim 20-30\%$ of the IRAS/DIRBE
values.  As noted above, the IRAS/DIRBE values have a larger
uncertainty, and potential systematic, in the Galactic Plane. 
Next, in Section 3.2, we show that the inferred extinction
based on the Tully-Fisher luminosity is consistent with the
CMD and SED-derived estimates.

Using the SED-derived extinction for each galaxy, G\,1 and G\,2, 
corrected infrared photometry 
is summarized in 
Table~1, including the aperture parameters used to extract the photometry.   
Extinction corrections were carried out for the NIR and 
MIR photometry, except for the 24$\micron$ results
whose
photometric
measurement uncertainties were larger than the actual extinction
correction appropriate to this long wavelength.
Formal uncertainties in
the measured fluxes ranged between 2--6\% in the NIR, and
$\sim$10--15\% for the IRAC and MIPS measurements. The MIR uncertainty is
dominated by the absolute calibration and the foreground stellar
contamination, both arise from the significant scattering of stellar light across the IRAC
focal planes.  Due to the heavy
foreground contamination from faint stars that we were unable to
identify and remove, systematic over-estimation of integrated fluxes
cannot be discounted at the 5\% level.
Fig.~2 shows the extinction-corrected SED for both galaxies.  For
clarity, we only show the GRASIL model template fit to G\,1, but note that G\,2 
also has a shape that fits the "old galaxy" template in the NIR portion of the SED.
For comparison we also show the MIR model spectrum of a dusty Sc-type
galaxy.  In the MIR, the galaxy emission significantly deviates from
the Rayleigh-Jeans tail due to the presence of strong PAH emission at
6.2 and 7.7 $\micron$, boosting the IRAC 5.8 and 8.0 $\micron$ bands.
The strong PAH lines are consistent with active or ongoing star formation.
Thermal dust continuum is also very strong in these galaxies as traced
by the 24 $\micron$ light, consistent with a large reservoir of gas
to fuel star formation (see below).
The combined NIR+MIR images reveal
G\,1 to be an edge-on disk galaxy with a weak nucleus/bulge, akin 
to Sc/Sd types, while G\,2 has a more face-on spiral orientation 
with the hint of a large-scale bar structure, possibly of SBbc-type. 

\subsection{Atomic Hydrogen}

A close-up view of the  ATCA \HI\ distribution map that is centered on the
GLIMPSE galaxies
is presented in Fig.~3.
Both galaxies are clearly detected in \HI. After primary-beam correction we 
measure \HI\ flux densities of \FHI\ = 1.1 Jy\kms\ and 5.6 Jy\kms\ 
for G\,1 and G\,2, respectively. Both galaxies have systemic 
velocities around 4500\kms\ (see insets in Fig.~3) which in the Local 
Group frame corresponds to $\sim 4320$\kms, giving a distance of 61 Mpc (\Ho 
= 72 km s$^{-1}$Mpc$^{-1}$) or a distance modulus of 33.93 mag. 
At this distance the \HI\ flux densities 
correspond to \HI\ masses of $10^9$\Msun\ and $5 \times 10^9$\Msun, 
i.e. typical values for star-forming spiral galaxies. 
The \HI\ properties are listed in Table~2.

The mean \HI\ velocity fields of both galaxies (not shown) indicate systematic
gradients, consistent with regularly rotating disks. Due to the low 
angular resolution, the position angle ($PA$) of the disks are difficult to
determine. For the galaxy G\,1 (top) we measure $\sim$140\degr.
Its large velocity width, \wtw\ = 429\kms, is consistent with the edge-on 
orientation seen in the infrared images (Fig.~1). Interestingly, 
the \HI\ profile is quite 
lopsided with the red-shifted horn being considerably stronger. This is 
also reflected in the \HI\ distribution which shows an extension of 
red-shifted velocity gas towards the south-east. A higher resolution \HI\ 
peak flux distribution (not shown) also shows the flux maximum shifted 
towards the south-east of the galaxy. Overall, the \HI\ distribution of
G\,1 appears slightly less extended than that of G\,2.
The $PA$ of the G\,2 disk is approximately --70\degr, compared to 
--40\degr\ as viewed in the infrared. The large velocity width, \wtw\ 
= 418\kms, is surprising given its rather face-on morphology on the 
infrared images, thus implying that the inclination-corrected linewidth
is very large indeed.  Finally, like G\,1, the  \HI\ spectrum of G\,2 appears to be asymetric.
In the higher resolution \HI\ peak flux distribution,  
the \HI\ maximum towards the southeastern part of the galaxy (blue-shifted 
velocities) is much stronger than that seen in the northwest (red-shifted 
velocities).

With its favorable disk orientation, well determined rotation velocity and near-infrared flux,
the absolute magnitude of galaxy G\,1 can be estimated using the near-infrared
Tully-Fisher (TF) relation.  We employ the K-band TF relation constructed by
Macri (2001), which correlates $M_K$ with the inclination corrected \HI\ linewidth
at the $20\%$ level, calibrated with K-band isophotal integrated fluxes determined
at the 21 mag/arcsec$^2$ level:  

\begin{math}
M_K = -22.53 - \displaystyle\biggl[10.0 \times [ log_{10}(w_{20}^i) - 2.5]\biggr]
\end{math}

The observed \wtw\ linewidth is 429\kms.
The K-band axis ratio of G\,1 is 0.25, corresponding to a inclination angle of 81 degrees,
assuming an intrinsic axis ratio of 0.2;  hence, $1 / sin(i)$ correction to the linewidth is very small,
$\sim 1\%$. Using a corrected linewidth $w_{20}^i$ = 434\kms, 
the inferred absolute magnitude using the above TF relation is -23.9 mag,
a luminosity that is $\sim 2.5\times$ brighter than $L_*$ for the field disk galaxy population
(Kochanek et al. 2001).   We note that deriving $M_K$ using other near-infrared TF relations
(e.g., Conselice et al. 2005) gives results that are in agreement to better than $\sim 10\%$. 
The redshift-based distance modulus is 33.9 mag (see above), which implies
that the apparent K-band magnitude for G\,1 should be about 10.0 mag.  
The 1.7 mag gap between this inferred apparent
magnitude and the actual apparent magnitude, 11.72 mag (see section 3.1), 
represents the total extinction of Milky Way foreground plus
G\,1 internal.  The resulting total extinction, Av = 15.1 mag, is remarkably
close to the extinction deduced from FIR dust emission, 
colors of field giant stars and of G\,1 itself (see section 3.1).
The clear implication is that
G\,1 follows the TF relation that describes disk galaxies whose primary 
baryonic mass and \HI\ components are intrinsic in origin. 

Following the same exercise with G\,2 is more problematic due
to the relatively face-on disk inclination.  The axis ratio implies an inclination that is
60 degree, or a $1 / sin(i)$ correction that is $\sim 15\%$, leading to an inclination-corrected
linewidth of 483\kms, and a TF luminosity of -24.4 mag.
The inferred apparent K-band brightness is 9.6 mag, which in comparison to
the actual K-band integrated brightness, 12.10 mag, implies a total extinction of greater 20 mag 
at visual wavelengths.  This value is much larger than what is inferred using the
the FIR, CMD or G\,2 SED, which suggests that G\,2 does not follow the TF relation, having a
disk rotation that is much greater than what is consistent with its K-band flux.

Taken as a whole, the large velocity width of G\,2 and the 
lopsided \HI\ distribution of both G\,1 and G\,2 
(Fig.~3) may be hinting at a tidal distorsion or 
a large-scale pressurized environment that is 
radially driving the \HI\ gas from symmetry. Although the Spitzer
IRSF and ATCA observations do not directly reveal any massive galaxies 
in the vicinity,
we cannot discount the presence of
\HI-poor or old-star dominated galaxies along this line-of-sight
due to 
stellar confusion and the
powerful
nebular emission arising from the Milky Way (see e.g., the northern portion of
Fig.~1).  In any event, the intriguing \HI\ line profiles 
merit further investigation.

\section{Discussion \& Summary}

To summarize, we have detected several new galaxy candidates in the GA region
using Spitzer MIR imaging from the GLIMPSE and MIPSGAL projects.  Two of these
sources were followed-up with deep NIR imaging and \HI\ line data:
(1) both are relatively luminous star-forming galaxies rich with gas and dust, 
and (2) 
based on their \HI\ radial velocities,
$\sim$4500\kms, are located within the nominal confines
of the \GA.
Previous Xray, optical/NIR and radio surveys of the GA did not detect these ZoA
galaxies, and there is reason to believe that many more galaxies
await discovery.

The region is extremely confused with 
gas, dust and stars, 
limiting the short-wavelength surveys, while
Galactic hydrogen emission
hampers traditional single-dish blind surveys of nearby galaxies
($<$500\kms) 
and the increase 
in Galactic radio continuum sources at lowest Galactic latitudes raises the
noise level and reduces the effectiveness of 21-cm surveys. 
Galactic synchrotron
blinds searches for 
extragalactic radio continuum sources, and Xray surveys are 
ineffective when the HI column density exceeds $\sim$30-50 $\times 10^{20}$\cm
(Ebeling et al. 2002).
Even with the Spitzer MIR imaging, we have found only a handful of
galaxies peering through the formidable stellar/dust mask, although
our follow-up multi-dish radio observations have proven to be very effective 
at penetrating the Plane.
Based on the TF-derived luminosities, both GLIMPSE galaxies are brighter
than typical field disk galaxies, so they may represent the upper tip of the 
luminosity function for field galaxies.
Both have strong IR emission
from star formation that does facilitate discrimination from the
foreground stars using the NIR-to-MIR colors.  The inferred foreground
extinction, $\sim$13-15 mag, pushes the photometry near the detection limit for disk
galaxies located at the distance of the Norma cluster; consequently,
most of the region studied is probably too opaque or confused with
bright nebular emission to detect most background galaxies using the
relatively shallow GLIMPSE data or the larger beam MIPSGAL data.
Deeper Spitzer imaging complemented with multi-dish radio \HI\ observations
are what is needed to uncover fainter galaxies or sources that are cloaked
by foreground stars and dust.

Are the GLIMPSE galaxies tracing a larger concentration of galaxies that bridge the
southern and northern hemispheres?  
Fig. 4 depicts the emerging view
of the GA and surrounding region, showing redshift-binned galaxy detections
from X-ray, optical, near-infrared and radio surveys.
The GLIMPSE galaxies appear in the middle panel. 
At low redshifts, $V_{hel} <
3500$\kms (top panel), the less-massive Centaurus Wall appears to extend from
the southern to the northern hemispheres at a slight angle with
respect to the supergalactic plane.  Whereas, at higher redshifts, $3500 -
6500$\kms, the combined filaments of the Pavo-Indus supercluster
(332\degr, $-24$\degr, 4200\kms) and the Norma cluster (325\degr,
$-7$\degr, 4900\kms) in the south appear to pass through the GLIMPSE
galaxies (317\degr, $-0.5$\degr, 4500\kms), bending to lower
longitudes and higher velocities, joining up with the Cen-Crux cluster
(306\degr, $+5.5$\degr, 5700--6200\kms) and
CIZA J1324.7-5736 (307.4\degr $+5.0$\degr, 5700\kms) to the north. 

The emerging 
portrait seems to be a
continuous "great wall"
that stretches from Pavo-Indus in the south ($\ell \sim 340\degr$), through
Norma and upwards (behind the Plane) toward Centaurus,
finally bending over to Vela in the north ($\ell \sim 270\degr$), thus extending
over $\sim$50\degr\ to 60\degr\ on the sky.  
This astonishingly large structure is analogous to the Virgo-Coma
"great wall" of galaxies discovered by the pioneering CfA redshift
survey of the northern hemisphere (Geller \& Huchra 1989).
Such an
extended mass concentration would surely
perturb the Hubble flow of galaxies in the local universe ($V_{hel} <
8,000$\kms), deserving its "great" attractor status as
surmised from large-scale flow studies of the last two decades  
(e.g.,
Kolatt, Dekel \& Lahav 1995).
And indeed, recent all-sky coverage studies of the galaxy 
density field using
x-ray selected (Kocevski \& Ebeling 2006) and near-infrared
selected (Erdogdu et al 2006a,b) galaxy samples clearly show the importance
of the GA to the local velocity field.  The GA is not the only attractor in the local universe,
for example the Perseus-Pisces Supercluster is another important contributer to
flow field, but it is the primary component within 8,000\kms.  At higher
velocities, the Shapley Concentration at $\sim$14,000\kms
is undoubtedly the most important structure in the local universe
(Hudson et al. 2004; Kocevski \& Ebeling 2006), although redshift and peculiar velocity studies
are highly incomplete at these large distances.    
The density field studies implicate the GA as the primary perturbing force 
to the Local Group and the larger Local Supercluster,
yet
the detailed morphology of the GA remains to be understood.
Combined mid-infrared and radio surveys of the ZoA hold great promise in
disentangling this cosmic web and delineating the Norma Wall from the foreground
Milky Way.

Future work will include similarly targeted observations of new GLIMPSE and MIPSGAL galaxy candidates.
Underway is a 
large area,
deep NIR survey with the IRSF, concentrating on the regions of the GA
that have \AV $<$ 10 mag, complemented by more sensitive interferometric radio \HI\
surveys that are optimized to finding galaxies in the velocity range
of the Norma Wall.  
Finally, we will propose deeper Spitzer MIR
imaging to probe the darkest regions of the ZoA.

\acknowledgments

We thank Sean Carey (SSC) for early access to the
rich MIPSGAL data set,
while
Lucas Macri (NOAO) and Karen Masters (Cornell) were a valuable
Tully-Fisher resource.
The contributions
of the whole Multibeam \HI\ ZOA Team are gratefully acknowledged, as
well as of K. Wakamatsu and T. Nagayama of the IRSF survey team of
the Norma Wall. RKK and PAW thank the NRF for financial support.
The Australia Telescope Compact Array is part of the Australia
Telescope which is funded by the Commonwealth of Australia for
operation as a National Facility managed by CSIRO.
The Spitzer Space Telescope is operated by the Jet Propulsion
Laboratory, California Institute of Technology, under contract with
NASA.  This research has made use of the NASA Extragalactic Database,
operated by IPAC/Caltech, under contract with NASA. 






\clearpage

\begin{figure} 
\includegraphics[angle=0,scale=0.9]{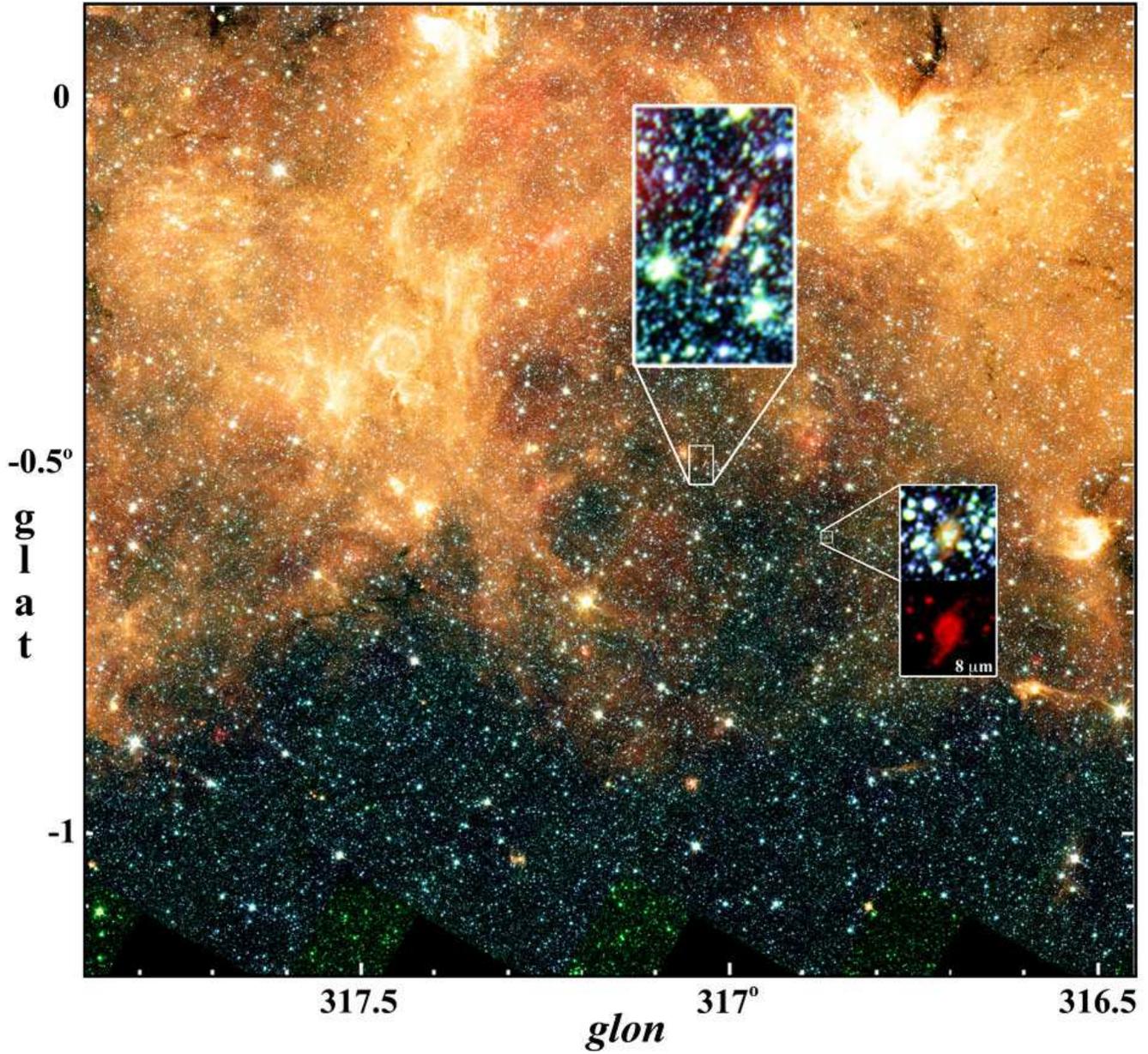}
\caption{Infrared view of the heart of the \GA\ region. This 
  color composite of Spitzer IRAC  (blue=$3.6 \micron$, green=$4.5 \micron$, 
  yellow=$5.8 \micron$, orange = $8 \micron$) 
and MIPS $24 \micron$ (deep red)
demonstrates the thick 
  veiling due to stars, gas and dust of the Milky Way. Nevertheless, 
  the MIR penetrates deeply enough to reveal at least two galaxies that 
  are shining through $\sim$15 magnitudes of visual extinction. The inset 
  images are constructed from deep IRSF $JHK_s (1-2.2 \micron$) images 
  combined with the MIR images.  Note the extremely red colors of the 
  edge-on disk galaxy, G\,1 (center), and the nearly face-on spiral 
  galaxy, G\,2 (lower right), due to extreme dust reddening and the 
  presence of strong PAH emission at $8 \micron$ (lower insert).  The 
  \HI\ redshift is $\sim$4500\kms\ for both galaxies, placing them 
  squarely in the Norma Wall of the \GA.}
\label{fig1}
\end{figure}

\clearpage

\begin{figure} 
\includegraphics[angle=270,scale=.65]{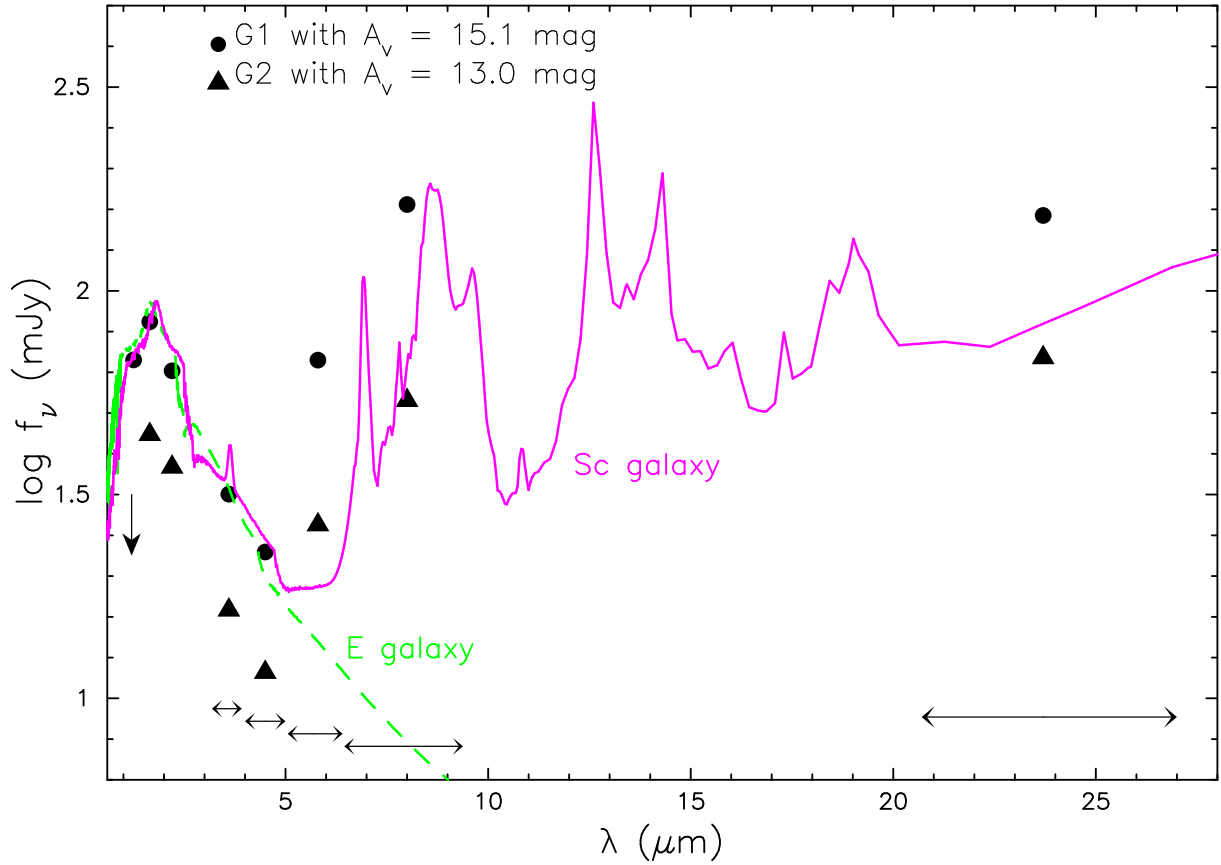}
\caption{Infrared spectral energy distribution, $f(\lambda)$, of the 
  galaxies G\,1 (filled circles) and G\,2 (filled triangles).  
  The flux density for each measurement has been corrected for selective 
  extinction (\AV\ = 15.1 and 13.0 mag, respectively).  For illustration, 
  two SED templates are fit to G\,1: (a) an old 13 Gyr E-type galaxy (green dashed line) 
  whose light is dominated by stellar photospheric emission from K/M giants, 
  and (b) a Sc-type galaxy (magenta solid line) whose MIR light is dominated by strong PAH 
  emission bands.  The broad IRAC and MIPS bandpasses are indicated at the bottom.}
\end{figure}

%

\clearpage

\begin{figure} 
\includegraphics[angle=0,scale=0.9]{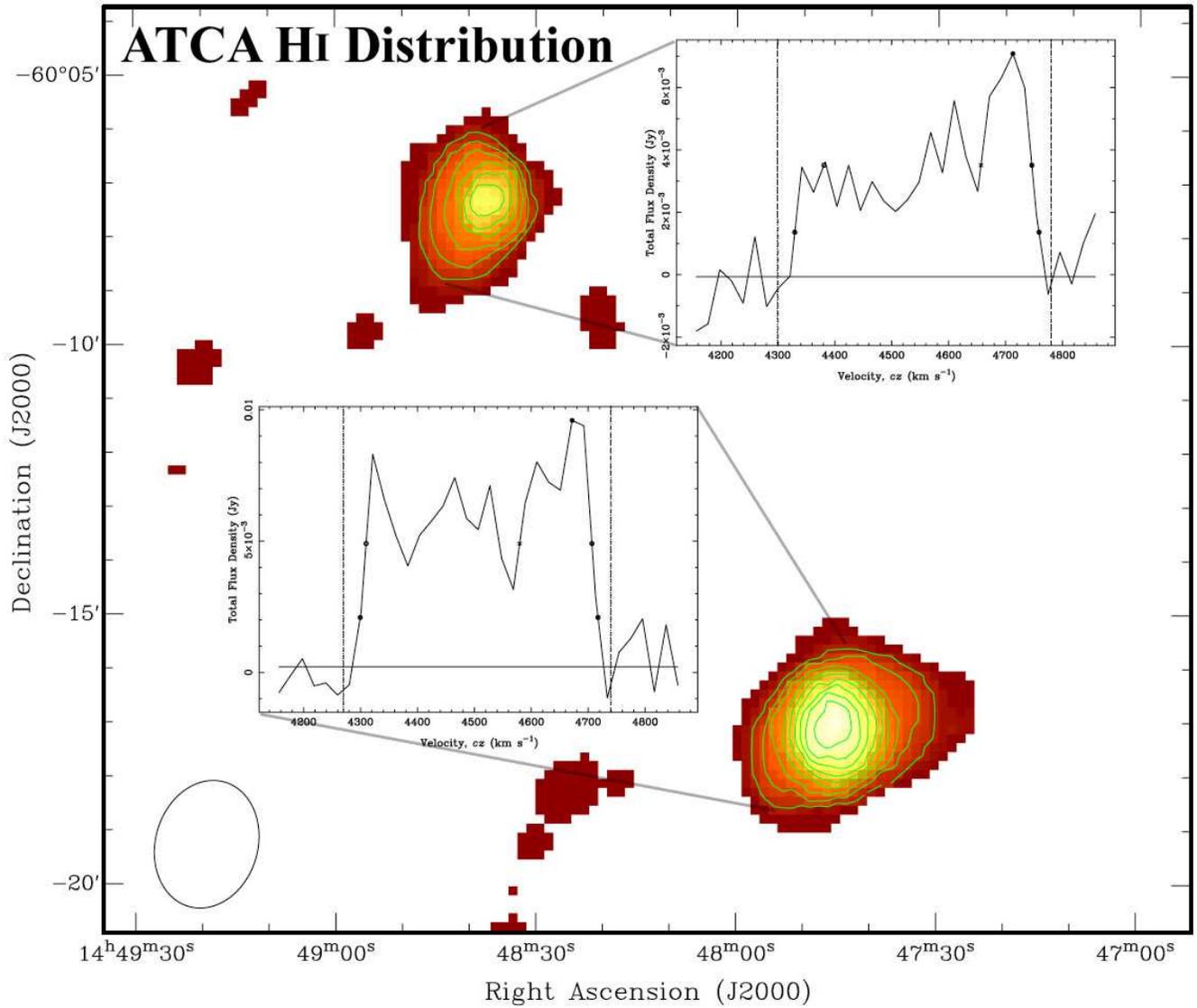}
\caption{ATCA \HI\ distribution map (moment 0) for the newly discovered 
  galaxies G\,1 (top) and G\,2 (bottom), before primary beam 
  correction. The synthesized beam ($144\arcsec \times 114\arcsec$; 
  `natural' weighting) is displayed at the bottom left. The two insets 
  show the corresponding integrated \HI\ spectra.
  The two galaxies are separated by $\sim$12\arcmin\ or $\sim$200 kpc 
  (assuming a distance of $D$ = 61 Mpc).}
\end{figure} 

\begin{figure} 
\includegraphics[angle=0,scale=0.85]{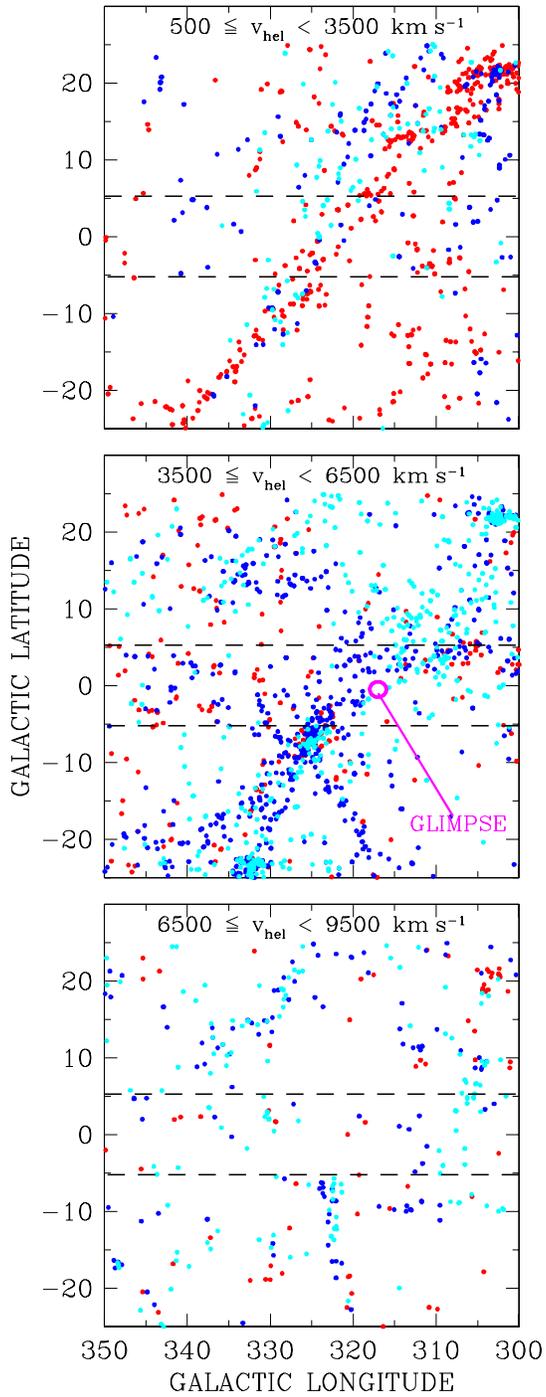}
\caption{Wide-field depiction of galaxies in the \GA. The three panels show galaxies
  with published radial velocities between 500 and 9500\kms, including
  data from the preliminary catalog of galaxies detected with the
  Multibeam ZOA \HI\ survey within $|b| \le 5\degr$ (dashed lines;
  Kraan-Korteweg 2005), separated into bins
  that reveal the nearby Centaurus filament (top panel) and the more distant Pavo/Norma
  Wall (middle panel). In each panel the lowest velocities are denoted 
  with cyan, moderate velocities with blue and the highest velocities 
  with red.  The location of the GLIMPSE galaxies is denoted with a 
  magenta circle in the middle panel.}
\end{figure}








\clearpage

\begin{deluxetable}{lccccccc} 
\tablecolumns{7}
\tabletypesize{\scriptsize}
\tablecaption{Infrared Photometry of the galaxies G\,1 and G\,2}
\tablewidth{0pt}
\tablehead{
\colhead{} & \multicolumn{3}{c}{GLIMPSE G\,1: 317.04\degr $-0.50$\degr}
           & \multicolumn{3}{c}{GLIMPSE G\,2: 316.87\degr $-0.60$\degr} \\
\colhead{} & \multicolumn{3}{c}{$a = 33.6$\arcsec, $b/a = 0.25$, 
              $\phi = -50$\degr, \AV = 15.1 mag}
           & \multicolumn{3}{c}{$a = 20.4$\arcsec, $b/a = 0.55$, 
              $\phi = -43$\degr, \AV = 13.0 mag} \\
\cline{2-4} \cline{5-7} \\
$\lambda$   & integrated flux & flux density & $\nu$L$_\nu$ 
            & integrated flux & flux density & $\nu$L$_\nu$  \\ 
$\micron$   & mag & mJy & $10^9$ \LLsun 
            & mag & mJy & $10^9$ \LLsun
}
\startdata
1.25 & 10.80$\pm$0.05 &  76.3$\pm$0.1 & 20.4  & $ <18$          & $<0.1$       &  0.02 \\
1.64 & 10.08$\pm$0.03 &  94.7$\pm$0.3 & 19.8  & 10.67$\pm$0.04  & 55.4$\pm$0.3 & 11.6  \\
2.15 &  9.91$\pm$0.02 &  72.1$\pm$0.7 & 11.2  & 10.54$\pm$0.02  & 40.5$\pm$0.5 &  6.29  \\
3.6  &  9.76          &  35.0         &  3.25 & 10.47           & 18.3         & 1.70   \\
4.5  &  9.68          &  24.1         &  1.79 & 10.38           & 12.7         &  0.94 \\
5.8  &  8.01          &  71.8         &  4.13 &  8.99           &  29.1        &  1.68  \\
8.0  &  6.41          & 175.7         &  7.33 &  7.66           &  55.6        &  2.32 \\
23.7 &  4.20          & 153.2         &  2.13  &  5.07           &  68.5        &  0.95  \\
\enddata
\tablecomments{Except for 24~$\micron$, all magnitudes and flux densities have been corrected for
  Milky Way dust extinction and internal extinction. The errors account for the photometry only, 
  not to the applied foreground extinction corrections. IRAC and MIPS measurements 
  have been aperture corrected using the extended source prescription 
  given by the Spitzer Science Center; typical uncertainties are $\sim$10\%. 
  The redshift-derived distance for both galaxies is 61 Mpc.}
\end{deluxetable}

\begin{table} 
\caption{\HI\ properties}
\label{tab:table2}
\begin{tabular}{lccccccc}
\hline
Galaxy Name & \HI\ max position& \vhel & \wfi & \wtw & \FHI & \MHI & M$_K$(TF)        \\
      &$\alpha,\delta$(J2000)  & \kkms & \kkms&\kkms &Jy\kms& $10^9$\Msun & mag \\
\hline
GLIMPSE G\,1 & $14^{\rm h}48^{\rm m}39^{\rm s},-60\degr07\arcmin20\arcsec$ 
             & 4563 & 365  & 429  & 1.1 & 1 & -23.9 \\
GLIMPSE G\,2 & $14^{\rm h}47^{\rm m}45^{\rm s},-60\degr17\arcmin09\arcsec$ 
             & 4508 & 397  & 418  & 5.6 & 5  & -24.4\\
\hline
\end{tabular}
\end{table}


\begin{thebibliography}{}
\bibitem{abe89}
Abell, G. O., Corwin, H. G. Jr., \& Olowin, R. P. 1989, \apjs, 70, 1 
\bibitem{arc99}
Arce, H. G. \& Goodman, A.A.  1999, ApJ, 512, L135.
\bibitem[Benjamin et al.(2003)]{ben03} 
Benjamin, R., et al. 2003, \pasp, 115, 953
\bibitem[Bonatto et al.(2004)]{bonatto04} Bonatto, C., Bica, E., \&
Girardi, L.\ 2004, \aap, 415, 571
\bibitem[Conselice et al.(2005))]{con2005}
Conselice, C.J., Bundy, K., Ellis, R.S., Brichmann, J., Vogt, N, \&
Phillips, A.  2005, \apj, 628, 160
\bibitem[Burstein et al.(1990)]{bur90} 
Burstein, D., Faber, S., \& Dressler, A.  1990, \apj, 354, 18
\bibitem[Carey 2006]{carey}
Carey, S. et al 2006, in preparation.
\bibitem[Draine 2004]{dr2003}
Draine, B.T.  2003, \araa, 41, 241
\bibitem[Ebeling et al. 2002]{eb02}
Ebeling, H., Mullis, C., Tully, R.B.  2002, \apj, 580, 774
\bibitem[Erdogdu et al.(2006a)]{er06a} 
Erdogdu, P., Huchra, J., Lahav, O., et al. 2006a, \mnras, 368, 151
\bibitem[Erdogdu et al.(2006b)]{er06b} 
Erdogdu, P., Lahav, O., Huchra, J., Colless, M., \& Jarrett, T.H.  
  2006b, \mnras, in press.
\bibitem[Ferraro et al.(2006)]{ferraro-method} 
Ferraro, F.R., Valenti, E., \& Origlia, L. 2006 (astro-ph/0605492)
\bibitem[Geller \& Huchra(1989)]{gh89}
Geller, M. \& Huchra, J. 1989, Science, 246, 897
\bibitem[Hudson et al.(2004)]{hud04} 
Hudson, M.J., Smith, R., Lucey, J., \& Branchini, E. 2004, \mnras, 352, 6
\bibitem[Jarrett(2000)]{jar00} 
Jarrett, T.H. 2000, \pasp, 112, 1008
\bibitem[Jarrett(2004)]{jar04}
Jarrett, T.H. 2004, PASA, 21, 396
\bibitem[Kocevski \& Ebeling(2006)]{koe06}
Kocevski, D.D. \& Ebeling, H.  2006, ApJ, in press; (astro-ph/0510106)
\bibitem[Kochanek et al (2001)]{koch01}
Kochanek, C. S., Pahre, M. A., Falco, E. E., Huchra, J. P., Mader, J., 
Jarrett, T. H., Chester, T., Cutri, R., Schneider, S.E., \apj, 560, 566
\bibitem[Kolatt, Dekel \& Lahav(1995)]{kol95} 
Kolatt, T., Dekel, A., \& Lahav, O.  1995, \mnras, 275, 797
\bibitem{kk05}
Kraan-Korteweg, R.C. 2005, RvMA, 18, 48 (astro-ph/0502217)
\bibitem{kk00}
Kraan-Korteweg, R.C., \& Lahav, O. 2000, A\&ARv, 10, 211
\bibitem{kk94}
Kraan-Korteweg, R.C., Cayatte, V., Balkowski, C., Fairall, A.P., \&
  Henning, P.A., 1994, ASP Conf. Ser. 67, 99
\bibitem[Kraan-Korteweg et al.(1996)]{kk96} 
Kraan-Korteweg, R.C., Woudt, P.A., Cayatte, V., Fairall, A.P., 
  Balkowski, C., \& Henning, P.A.  1996, Nature, 379, 519 
\bibitem{lahav87}
Lahav, O. 1987, \mnras, 225, 213
\bibitem[L{\'o}pez-Corredoira et al.(2005)]{lopez-bulge} 
L{\'o}pez-Corredoira, M., Cabrera-Lavers, A., \& Gerhard, O.~E.\ 2005, 
\aap, 439, 107 
\bibitem[Lynden-Bell et al.(1988)]{lbell88} 
Lynden-Bell, D., Faber, S.M., Burstein, D., Davies, R.L., Dressler, A.,
  Terlevich, R., \& Wegner, G. 1988, \apj, 326, 19
\bibitem{lbell89}
Lynden-Bell, D., Lahav, O., \& Burstein, D. 1989, \mnras, 241, 325
\bibitem[Macri(2001]{macri01}
Macri, L.M.  2001, PhD thesis, Harvard University.
\bibitem{masters03}
Masters, K.L., Giovanelli, R. \& Haynes, M.P.  2003, \aj, 126, 158.
\bibitem[Nagata \& Glass(2000)]{ng00}  
Nagata, T., \& Glass, I.S.  2000, MNASSA, 59, 110
\bibitem{nag03}
Nagayama, T., Nagashima, C., Nakajima, Y., et al. 2003, Proc.SPIE, 4841, 459
\bibitem{nag04}
Nagayama, T., Woudt, P.A., Nagashima, C., et al. 2004, \mnras, 354, 980 
\bibitem[Nishiyama et al.(2006)]{nishayama-galcen} Nishiyama, S., et 
al.\ 2006, \apj, 647, 1093 
\bibitem{rad06}
Radburn, D.J., Lucey, J.R., Woudt, P.A., Kraan-Korteweg, R.C., \& 
  Watson, F.G. 2006, \mnras, 369, 117.
\bibitem[Rocha-Pinto et al.(2006)]{helio} Rocha-Pinto, H.~J., 
Majewski, S.~R., Skrutskie, M.~F., Patterson, R.~J., Nakanishi, H., 
Mu{\~n}oz, R.~R., \& Sofue, Y.\ 2006, \apjl, 640, L147
\bibitem[Salaris \& Girardi(2002)]{salaris02} Salaris, M., \& 
Girardi, L.\ 2002, \mnras, 337, 332 
\bibitem[Salaris \& Girardi(2005)]{salaris-trbg} Salaris, M., \& 
Girardi, L.\ 2005, \mnras, 357, 669
\bibitem[Schlegel et al.(1998)]{schlegel} 
Schlegel D.J., Finkbeiner D.P., \& Davis M. 1998, \apj, 500, 525
\bibitem[Silva et al.(1998)]{silva98} 
Silva, L., Granato, G., Bressan, A., \& Danese, L.  1998, \apj, 509, 103
\bibitem{}
Schr\"oder, A., Mamon, G.A., Kraan-Korteweg, R.C., \& Woudt, P.A. 2006,
  A\&A, submitted (astro-ph/0607108)
\bibitem[Woudt \& Kraan-Korteweg(2001)]{wk01} 
Woudt, P.A., \& Kraan-Korteweg, R.C.  2001, A\&A, 380, 441
\bibitem {wkf99}
Woudt P.A., Kraan-Korteweg, R.C., \& Fairall, A.P. 1999, A\&A, 352, 39
\end{thebibliography}
\end{document}